\documentclass[aps,prb,twocolumn,showpacs,amsmath,floatfix]{revtex4}
\usepackage{graphicx}
\usepackage{dcolumn}

\begin{document}


\title{Hybrid density functional calculations of the band gap of Ga$_x$In$_{1-x}$N}

\author{Xifan Wu$^{1}$, Eric J. Walter$^{2}$, Andrew M. Rappe$^{3}$, 
Roberto Car$^{1}$, and Annabella Selloni$^{1}$}

\address{$^{1}$Chemistry Department, Princeton University, Princeton, NJ
08544-0001,USA}

\address{$^{2}$Department of Physics, College of William and Mary, 
Williamsburg, Virginia 23187-8795, USA}

\address{$^{3}$The Makineni Theoretical Laboratories, Department of Chemistry, 
University of Pennsylvania, Philadelphia, Pennsylvania 19104-6323, USA}

\date{\today}

\begin{abstract}
Recent theoretical work has provided evidence that hybrid functionals, which include 
a fraction of exact (Hartree Fock) exchange in the density functional theory (DFT) 
exchange and correlation terms, significantly improve the description of band gaps 
of semiconductors compared with local and  semilocal approximations. Based on a 
recently developed order-$N$ method for calculating the exact exchange in extended 
insulating systems, we have implemented an efficient scheme to determine the hybrid 
functional band gap. We use this scheme to study the band gap and other electronic 
properties of the ternary compound In$_{1-x}$Ga$_{x}$N using a 64-atom supercell model.

\end{abstract}


\maketitle


\marginparwidth 2.7in

\marginparsep 0.5in

\def\xwm#1{\marginpar{\small XW: #1}}
\def\rcm#1{\marginpar{\small RC: #1}}
\def\asm#1{\marginpar{\small AS: #1}}

The design of novel functional semiconductors with given values of the 
energy band gap is an area of intense research
\cite{Zunger_inverse, Vasp_alloy, ZnO_GaN, Wang, Bennett}.  In
particular, much attention is focused on the band gap engineering of
group-III nitride semiconductors, whose remarkable optical properties
are important for optoelectronic device applications
\cite{InGaN_review, Zunger_InGaN}.  To guide the search for compounds
with tailored properties \cite{Zunger_inverse}, experimental studies
are often accompanied by electronic structure calculations based on
density functional theory (DFT) \cite{DFT}.  For these calculations,
the local-density(LDA) or generalized gradient approximations(GGA) are
typically used.  Due to the delocalization error of the LDA and GGA
exchange and correlation functionals, however, these approaches
severely underestimate the materials band gaps \cite{Yang_Science,
Yang_PRB}.

As shown by several recent studies \cite{HSE_review}, 
a significant improvement in the
description of semiconductor and insulator band gaps is generally
obtained by using hybrid functionals \cite{Hybrid}, in which some exact
(Hartree-Fock) exchange is mixed into the exchange and correlation
functional. This reduces the delocalization and derivative
discontinuity errors of (semi)local functionals \cite{Yang_Science,
Yang_PRB, Vasp_alloy, HSE_review}.  However, because of the
considerable computational cost of evaluating the non-local exact
exchange term, hybrid functionals have been mostly applied to systems
with small unit cells \cite{PBE0_vasp}.  For the modeling of systems
where a large supercell is needed, an additional screened exchange
approximation is usually made to relieve the computational burden
\cite{Vasp_alloy, HSE_review}.

Recently Wu {\it et al.} (WSC) \cite {Xifan} introduced an order-$N$
method to calculate the exact exchange in extended insulating
systems. The WSC method is based on a localized Wannier function
representation of the occupied (valence) space, so that the exchange
interaction between two orbitals decays rapidly with the distance
between their centers. A truncation can thus be introduced, which
greatly reduces the computational cost. The effectiveness of the WSC
method was demonstrated by ground state electronic minimizations for
crystalline silicon in supercells with 64 and 216 atoms.

In this paper, we extend the WSC scheme to compute hybrid functional
band gaps. To this end, the system's first (few) empty conduction
state(s) is(are) determined starting from the ground state calculated
via the WSC method. With hybrid functionals, this requires the
computation of the pair exchange between the empty state and each
valence orbital.  Even though the empty state is delocalized, the
product between this state and a valence orbital is well localized, 
so that the corresponding exchange interaction can be truncated 
as in the original WSC method \cite{Xifan}.  
We apply our scheme to determine
the band gap of In$_{1-x}$Ga$_{x}$N, a ternary nitride semiconductor
of great technological interest, and of its parent compounds, InN and
GaN, using the PBE0 hybrid functional \cite{PBE0}. Our results show
that, compared to the semi-local PBE functional, PBE0 gives a
considerably improved description of the band gap, as well as of the
cation $d$ state binding energy, which is also poorly decribed by the
semilocal functionals.

The PBE0 hybrid functional is constructed by mixing 25\% of exact
exchange with the GGA-PBE exchange \cite{PBE0}, while the correlation
potential is still represented by the corresponding functional in PBE
\cite{PBE},

\begin{equation}
E_{xc}^{\rm PBE0} = \frac{1}{4}E_x+\frac{3}{4}E_{x}^{\rm PBE} + E_{c}^{\rm PBE}.
\label{eq:PBE0_energy}
\end{equation}

Here $E_x$ denotes the exact exchange energy, $E_x^{\rm PBE}$ is the
PBE exchange, and $E_c^{\rm PBE}$ is the PBE correlation functional.
$E_x$ has the usual Hartree-Fock form in terms of one-electron
orbitals.
In the WSC method, this term is expressed in terms of localized
Wannier orbitals $\{\widetilde\varphi_i\}$.  These are obtained
through an unitary transformation of the delocalized Bloch states
$\{\varphi_i\}$ corresponding to occupied bands. In particular, we use
maximally localized Wannier functions (MLWFs) \cite{MLWF}, which are exponentially
localized.  In this way, a significant truncation in both number and
size of exchange pairs can be achieved in real space.

We now turn to the calculation of the band gap.  In extended
insulating systems the band gap is simply given by the difference
between the eigenvalue of the highest occupied and the lowest empty
state. Once the ground state has been minimized self-consistently, the
eigenvalue of the empty state $\varphi_e$ can be obtained through a
simple non-selfconsistent calculation. With the hybrid PBE0
functional, the equation for $\varphi_e$ is
\begin{eqnarray}
\Bigl (&-&\frac{1}{2}\nabla^2 + V_{\rm ion}({\bf r}) + V_{\rm H}[\,\rho^{\rm val}({\bf r}) \,] 
 + \frac{3}{4}V_x^{\rm PBE}[\,\rho^{\rm val}({\bf r}) \, ] \cr
 &+& V_c^{\rm PBE}[\,\rho^{\rm val}({\bf r}) \, ] 
 \Bigr ) \times  \varphi_{e}({\bf r}) 
+ \frac{1}{4}\int V_x^{\rm val}({\bf r, r'}) \varphi_{e}({\bf r'})d{\bf r'} \cr
&=& \varepsilon_e \varphi_{e}({\bf r}) ,
\label{eq:eigenvalue}
\end{eqnarray}
In the above expression we have assumed, for simplicity, a
closed-shell system with $N/2$ doubly occupied one-electron states
(extension to spin-polarized systems is straightforward); $V_{\rm H}$
and $V_{\rm ion}$ are the Hartree and the ionic (pseudo-)potentials,
respectively; $V_x^{\rm PBE}$ and $V_c^{\rm PBE}$ are the PBE exchange
and correlation potentials.  We note that $V_{\rm H}$, $V_x^{\rm PBE}$
and $V_c^{\rm PBE}$ are fixed operators as they only depend on the (fixed)
valence charge density $\rho^{\rm val}({\bf r}) = \sum_j^{\rm
  occ}\varphi_j^*({\bf r})\varphi_j({\bf r})$.  Finally, the non-local
exact exchange potential ${V}_x^{\rm val}({\bf r, r'})$ is given by:
\begin{equation}
{V}_x^{\rm val}({\bf r, r'})= -2 \sum_j^{\rm occ} 
\frac{\widetilde\varphi_j^*({\bf r'})\widetilde\varphi_j({\bf r})}{\vert {\bf r} - {\bf r'} \vert },
\label{eq:Exx_potential}
\end{equation}
where the sum runs over all the occupied states. This potential
describes the exchange interaction between the empty state and each of
the valence MLWFs $\{\widetilde\varphi_j \}$.

The action of ${V}_x^{\rm val}({\bf r, r'})$ on the empty 
state $\varphi_e$ in Eq.~(\ref{eq:Exx_potential}) is given by:
\begin{eqnarray}
D_x^e ({\bf r} ) &=& -2  \sum_j^{\rm occ} \int d {\bf r'}
\frac{ \widetilde\varphi_j ^* ( {\bf r'}) \varphi_e ({\bf r'}) }
{ \vert {\bf r} - {\bf r'} \vert } \times \widetilde\varphi_j ( {\bf r} )  \cr
&=&  -2  \sum_{j}^{\rm occ}v_{ej}({\bf r}) \widetilde\varphi_{j}({\bf r}) 
\label{eq:action}
\end{eqnarray}
Here $v_{ej}$ is the Coulomb potential originating from the ``exchange
charge'' $\rho_{\rm ej} = \widetilde\varphi_j ^* ( {\bf r'}) \varphi_e
({\bf r'}) $, and satisfies the Poisson equation:
\begin{equation}
\nabla^2 v_{ej}  =  - 4\pi\rho_{ej} 
\label{eq:Poisson}
\end{equation}
It is important to note that, while the empty eigenstate of Eq.~(\ref{eq:eigenvalue}) is Bloch like
and delocalized in real space, the exchange pair density $\rho_{ej}$
is confined by the valence MLWFs that are well localized in real
space.  As a result, the Poisson equation, Eq.~(\ref{eq:Poisson}), and
the action of the exchange operator, Eq.~(\ref{eq:action}), need only
be solved in the region where $\widetilde\varphi_j \neq 0$.

We have implemented the above computational procedure for calculating
the PBE0 band gap in the CP code of the Quantum-ESPRESSO
package.~\cite{QuantumEspresso} The procedure works as a post
processing feature following a PBE0 ground state calculation by the
MLWF-based WSC method.  In this work, we use it to calculate the
electronic structure, particularly the band gap, of GaN, InN, and
In$_{1-x}$Ga$_x$N in the zincblende phase.  These systems are
computationally challenging because InN and In-rich In$_{1-x}$Ga$_x$N
are incorrectly predicted to be metallic by standard GGA calculations.

The calculations were performed using a 64-atom cubic supercell to
model both In$_{1-x}$Ga$_x$N and its parent compounds, GaN and InN.
For each Ga concentration $x$ in the ternary In$_{1-x}$Ga$_x$N
compound, only a few selected atomic configurations were considered,
with no specific treatment of disorder effects, as e.g.  in Refs.~
\onlinecite{Zunger_InGaN, Wang}; within our limited sampling, a very
weak dependence of the calculated band gap on the specific cation
arrangement was observed.  For direct comparison with experiments and
other theoretical results, the experimental lattice constants of GaN
(a = 4.50 \AA) and InN (a = 4.98 \AA ) were used, while the lattice
parameter of the alloy was determined by linear interpolation.

\begin{table}[ht]
\caption{Pseudopotential generation parameters. Here ``ref.'' refers 
to the reference state occupation, r$_c$ refers to the cut-off radius, 
$q_c$ is the cut-off wavevector and $N_B$ is the number of Bessel
functions used for each channel (see Ref.~\onlinecite{RRKJ}).}
\begin{ruledtabular}
\begin{tabular}{ccddd}
Atom & parameter & \multicolumn{1}{c}{$s$} & \multicolumn{1}{c}{$p$} & \multicolumn{1}{c}{$d$} \\\hline
N & ref. & 2.0 & 3.0 & \multicolumn{1}{c}{--}\\
  & \multicolumn{1}{c}{${\rm r}_c$} & 1.30 & 1.30 & \multicolumn{1}{c}{--}\\
  & \multicolumn{1}{c}{$q_c$} & 7.50 & 7.50 & \multicolumn{1}{c}{--}\\
  & \multicolumn{1}{c}{$N_B$} & \multicolumn{1}{c}{10}& \multicolumn{1}{c}{10}& \multicolumn{1}{c}{--}\\ \hline
Ga & ref. & 2.0 & 1.0 & 10.0\\
  & \multicolumn{1}{c}{${\rm r}_c$} & 1.80 & 2.20 & 1.80\\
  & \multicolumn{1}{c}{$q_c$} & 8.00 & 8.00 & 8.36\\
  & \multicolumn{1}{c}{$N_B$} & \multicolumn{1}{c}{6}& \multicolumn{1}{c}{8}& \multicolumn{1}{c}{10}\\ \hline
In & ref. & 2.0 & 1.0 & 10.0\\
  & \multicolumn{1}{c}{${\rm r}_c$} & 1.90 & 2.30 & 1.80\\
  & \multicolumn{1}{c}{$q_c$} & 8.00 & 8.00 & 8.00\\
  & \multicolumn{1}{c}{$N_B$} & \multicolumn{1}{c}{8}& \multicolumn{1}{c}{8}& \multicolumn{1}{c}{8}\\ 
\end{tabular}
\end{ruledtabular}
\label{table:pspinfo}
\end{table}

Table~\ref{table:pspinfo} shows the reference states and cut-off radii
used to construct the pseudopotentials used in this study.  All
pseudopotentials were generated using the OPIUM code \cite{OPIUM}. 

Unlike with traditional density functional theory, 
Hartree-Fock pseudopotentials require extra care in their
construction. This arises from the non-local form of the Hartree-Fock
exchange potential \cite{trail_needs1,bk_exact_xc,stadele_exact_xc,engel_exact_xc}.
The presence of the non-local exchange potential in Hartree-Fock or Hartree-Fock/DFT
hybrids will often yield pseudopotentials with an unphysical, long-range
tail. A correction procedure is necessary to remove this tail and
restore the correct long-range behavior of the pseudopotential while
maintaining the eigenvalue spectrum and logarithmic
derivatives. Recent work \cite{trail_needs1, trail_needs2, AWR} has
shown that this approach yields highly accurate Hartre-Fock
pseudopotentials.

The pseudopotentials were norm-conserving/RRKJ type \cite{RRKJ} and
were generated from self-consistent PBE0 all-electron reference
states using the approach of Ref.~\onlinecite{AWR}. The Ga and In
pseudopotentials were obtained from scalar-relativistic solutions,
while the N pseudopotential was non-relativistic. The local potential
was the $s$ channel for all cases.  The semi-core $d$ electrons were
treated as valence electrons in In and Ga (this corresponds to 576
valence electrons, {\em i.e.} 288 occupied states, in the 64-atom
supercell). The plane-wave energy cutoff was 70 Ry and the Brillouin
zone was sampled at the $\Gamma$ point.  Atomic positions in the
supercell were relaxed at the GGA-PBE level.

The PBE0 ground state was determined by the WSC method, using MLWFs to
calculate the exchange interaction among valence electrons
\cite{Xifan}. While the MLWFs generated from the PBE ground state
often give an excellent initial guess for the PBE0 calculations, for
InN and In rich Ga$_x$In$_{1-x}$N alloy configurations, the PBE ground
state shows an incorrect ordering of the energy bands. For this
reason, instead of PBE Wannier orbitals we used a set of fictitious
localized orbitals at the guess bonding centers as the trial solutions
for Eq.~(\ref{eq:eigenvalue}). This procedure was essential to obtain
the PBE0 ground state with correct symmetry for InN and In rich
Ga$_x$In$_{1-x}$N.
In the empty state calculations, for each PBE0 ground state MLWF we
first defined an orthorhombic box
such that outside this box 
$\rho_{ej}({\bf r})$ is smaller than a given cut-off value $\rho^{\rm
  cut}$; we take this cut-off equal to $2 \times 10^{-4}\ {\rm
  bohr}^{-3}$ in the present work. Then Eq.~(\ref{eq:Poisson}) is
solved by the conjugate gradient method \cite{Xifan}, and for each
pair $\rho_{\rm ej}$ formed by the empty state and a PBE0 ground state
MLWF its action Eq.~(\ref{eq:action}) is applied only inside the above
truncated box. Finally with this $D_x^e({\bf r})$,
Eq.~(\ref{eq:eigenvalue}) is solved via a damped second order
Car-Parrinello dynamics \cite{RC review}.

\begin{figure}[ht]
\includegraphics[width=3.1in]{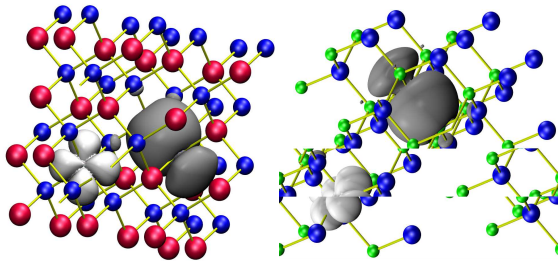}
\caption{\label{fig1} (Color online.) Isosurfaces of typical 
$d$-like and $sp^3$ like  Wannier orbitals in the InN (on the left)
and GaN (on the right) 64-atom supercell. The Ga, In 
and N atoms are denoted by the green, red and blue spheres respectively.}
\end{figure}

Representative MLWFs for InN in its PBE0 ground state are shown in
Fig.~\ref{fig1}. Two types of valence MLWFs are present in our
calculations, $d$-like Wannier orbitals centered at the In sites, and
covalent $sp^3$-like orbitals centered between the cations and the
anions. As one can see from the figure, the $d$-like orbitals
originating from the cation semi-core states are more localized than
the $sp^3$-like ones. The valence MLWFs are qualitatively similar for
GaN, except for a slightly more pronounced localization related to the
larger band gap.
\begin{table}[ht]
\caption{Valence band width, band gap and average $d$-band binding energy (eV) of GaN and InN.}
\begin{ruledtabular}
\begin{tabular}{llddddd}
& & \multicolumn{1}{c}{\rm  VBW} 
& \multicolumn{1}{c}{$E_g$ } & \multicolumn{1}{c}{$E_d$ }\\ 
\hline
GaN &PBE0-MLWFs  & 17.70 & 3.52 & -16.16 \\ 
& PBE & 16.14 & 1.60 & -13.62  \\
& PBE0, plane waves \tablenotemark[1] & 17.72  & 3.61 &   \\
& GW \tablenotemark[2] &   & 3.53 &  -16.5 \\
& Experiment \tablenotemark[3] &  & 3.3 &  -17.7 \\
\hline
InN &PBE0-MLWFs & 17.04 & 1.09 & -15.30  \\ 
& PBE results & 15.04 & -0.04 & -13.48   \\
& GW \tablenotemark[2] &   & 0.78 &  -15.3 \\
& Experiment  &  & 0.61 \tablenotemark[2] & -16.0 \tablenotemark[3] \\
\end{tabular}
\end{ruledtabular}
\label{table:Gap}
\tablenotetext[1] {Reciprocal space method in 
PWSCF (Ref. \onlinecite{QuantumEspresso})
in 2-atom cell and 4$\times$4$\times$4 $k$ points}
\tablenotetext[2] {Reference~\protect\onlinecite{InN_vasp}.} 
\tablenotetext[3] {Reference~\protect\onlinecite{InN_d}.}
\end{table}

The band structure properties of GaN and InN that result from our
PBE0-MLWFs calculations are summarized in Table~\ref{table:Gap}. Here
we report the valence band width (VBW), the band gap $E_g$ and the
average $d$-band binding energy $E_d$, and compare them to PBE
calculations (performed with the same 64-atom supercell used for the
PBE0 calculations) and experimental results. For further comparison,
we also report the results of PBE0 calculations performed using the
reciprocal space implementation in Ref.~\onlinecite{QuantumEspresso};
we can see that the agreement between these results and our MLWF-based
calculations is very good.  From Table~\ref{table:Gap} it appears that
the GGA-PBE results significantly overestimate the energetic position
of the cation $d$-bands.  Because of the $pd$ repulsion, the
overestimated $d$ bands level in turn pushes the $p$ band upwards,
resulting in an underestimated band gap.  For InN, this effect leads
to a wrong ordering of the $\Gamma_{1c}$ and $\Gamma_{15v}$ energy
levels, and thus to the incorrect prediction of a metallic ground
state.
In the PBE0 calculations, the inclusion of exact exchange reduces the
delocalization error. As shown by Table~\ref{table:Gap}, the PBE0 VBW
is larger and the $d$-bands level shifts downwards, in better
agreement with the experiment. In turn, this leads to a considerable
improvement of the band gaps of both InN and GaN with respect to
experiment; in particular, the PBE0 band gap becomes 1.09 eV for InN.
It is also worth noticing that calculation of the PBE0 band gap using 
a PBE pseudopotential yields a $\sim$ 0.2 eV smaller value than that
obtained with the PBE0 pseudopotential.

\begin{figure}[ht]
\includegraphics[width=2.8in]{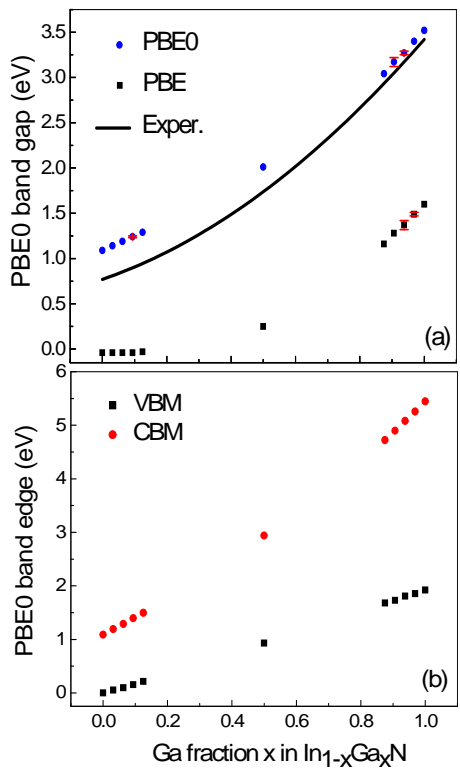}
\caption{\label{fig2} (Color online.) (a) PBE0, PBE and experimental 
band gap of dependence 
Ga fraction $x$ (b) Valence band maximum (VBM) and 
conduction band minimum as a function of Ga fraction $x$ in 
In$_{1-x}$Ga$_x$N }
\end{figure}

\begin{figure}[ht]
\includegraphics[width=4.0in]{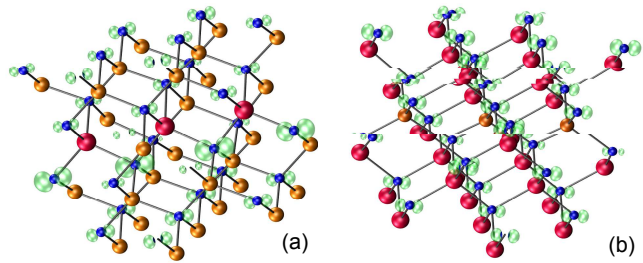}
\caption{\label{fig3} (Color online.) Isosurfaces of PBE0 eigenstate 
(a) In$_3$Ga$_{29}$N where 3 In atoms forms a zigzag chain
structure; (b) Ga$_3$In$_{29}$N where 3 Ga atoms forms a zigzag chain 
N atoms are denoted by red, orange and blue spheres respectively }
\end{figure}

Besides confirming the good performance of hybrid functionals for band
gap predictions, the above results for InN and GaN provide evidence of
the reliability of our procedure for calculating the PBE0 band gap.
We have thus applied this procedure to the study of the ternary
In$_{1-x}$Ga$_x$N compound, a system for which the standard reciprocal
space approach to calculate the exact exchange would be extremely
cumbersome. Instead, our order-$N$ scheme is well suited to treat
systems for which large supercells are needed.
Using a 64-atom supercell, we then considered In$_{1-x}$Ga$_x$N models
with 1(31), 2(30), 3(29), 4(28), 16(16), 28(4), 29(3), 30(2) and 31(1)
Ga(In) cations, which correspond to $x$ = 0.031, 0.063, 0.094, 0.125,
0.5, 0.875, 0.906, 0.938, and 0.969. For each value of $x$ and a given
configuration of Ga(In) atoms, the atomic positions were relaxed at
the PBE level.
The computed PBE0 band gap of In$_{1-x}$Ga$_x$N as a function of the
Ga fraction $x$ is shown in Fig.~\ref{fig2}(a), together with
experimental \cite{bowing} and PBE results. We can see that PBE not
only significantly underestimates the band gap but incorrectly shows a
metallic ground state for $x<0.5$.  By contrast, a direct band gap at
the $\Gamma$ point is found for all values of $x$ at the PBE0
level. Moreover, PBE0 predicts a large band gap bowing effect, in
qualitative agreement with the experiment \cite{bowing}. The band gap
can be fitted to the quadratic form
\begin{equation}
E_{g}^{\rm alloy} = xE_g^{\rm GaN}+(1-x)E_g^{\rm InN}-x(1-x)b
\label{eq:bowing}
\end{equation}
from which a bowing coefficient $b^{\rm PBE0}$ = 1.63 eV can be extracted,
similar to the value, 1.67 eV, found in previous screened-exchange
density functional ($sx$-LDA) calculations \cite{Wang}. However, this is
somewhat larger than the experimental value $b^{\rm expt}$ = 1.43 eV
\cite{bowing}, likely because of the overestimated PBE0 band gap for
the In-rich compounds.
To gain more insight into the origin of the large band gap bowing, we
have examined how the valence band maximum (VBM) and conduction band
minimum (CBM) depend separately on $x$, see Fig.~\ref{fig3}(a).  In
this analysis, the average electrostatic potential was taken as the
reference for the band alignment.  It can be seen that the VBM
increases almost linearly with $x$, whereas the CBM shows a stronger
nonlinear increase which is responsible for the large bowing
coefficient of the alloy.

The electronic states in proximity of the VBM are important for the
pholuminescence properties of In$_{1-x}$Ga$_x$N. These states have the
character of $p$ orbitals localized at the N sites. Previous
theoretical studies of In$_{1-x}$Ga$_x$N found that in Ga-rich alloys
the amplitude of these states is enhanced at N sites close to In
impurities\cite{Wang, Zunger_InGaN}, suggesting a localization of
photoexcited holes at such sites.  This interesting result is
confirmed by our PBE0 hybrid calculations.  The enhancement, or hole
localization, is particularly evident when the In impurities are
clustered to form a zigzag In-N-In-N-In chain, as shown in
Fig.~\ref{fig3}(a). This localization has been suggested to be the
reason of the high efficiency of In$_{1-x}$Ga$_x$N based emitting
devices \cite{Wang, InGaN_review}.  Interestingly, we found that there
is an opposite effect for the case of Ga impurities in In rich
alloys. Here, a reduction of the $p$ states at the N sites along the
Ga-N-Ga-N-Ga-N chain is observed, see Fig.~\ref{fig3}(b).

In conclusion, we have described an efficient procedure to calculate
the band gap of extended insulating systems using hybrid
functionals. This procedure is based on the recently developed WSC
order-$N$ method, in which the Hartree Fock exchange is calculated
using MLWFs, and can therefore be used to study the band gap and other
electronic properties of systems with large unit cells. We have
demonstrated the effectiveness of our approach by a study of the band
gap of a ternary compound, In$_{1-x}$Ga$_x$N, that we have modeled
using a 64-atom supercell. Hybrid functional results for this
important material are here reported for the first time, without the
approximation of screened exchange, and show a much better agreement
with experiment than conventional DFT-GGA or LDA calculations.  Our
approach can be widely used for the band gap engineering problem in
semiconductor alloys.

\acknowledgments
This work has been supported by the Department Of Energy under grant
DE-FG02-06ER-46344, grant DE-FG02-05ER46201 and by AFOSR-MURI  F49620-03-1-0330. 
A. M. R. was supported by the (US) Department of Energy under grant
DE-FG02-07ER46431


\end{document}